\documentclass[]{spie}  

\usepackage{amsmath,amsfonts,amssymb}
\usepackage{graphicx}
\usepackage[colorlinks=true, allcolors=blue]{hyperref}
\usepackage{authblk}

\usepackage{multicol}
\usepackage{graphicx}
\usepackage{gensymb}
\usepackage{caption}
\usepackage{subcaption}
\usepackage{adjustbox}
\usepackage{mathtools}
\usepackage{bm}
\usepackage{ctable} 
\usepackage{float} 
\usepackage{aas_macros}

\title{Auto-tuned thermal control on stratospheric balloon experiments}

\author[a,c]{Susan Redmond}
\author[e]{Steven J. Benton}
\author[d]{Anthony M. Brown}
\author[d]{Paul Clark}
\author[a]{Christopher J. Damaren}
\author[f,g]{Tim Eifler}
\author[e]{Aurelien A. Fraisse}
\author[b,c]{Mathew N. Galloway}
\author[b,c]{John W. Hartley}
\author[h]{Mathilde Jauzac}
\author[e]{William C. Jones}
\author[e]{Lun Li}
\author[e]{Thuy Vy T. Luu}
\author[d]{Richard J. Massey}
\author[f]{Jacqueline Mccleary}
\author[b,c,i,j]{C. Barth Netterfield}
\author[c,i]{Ivan L. Padilla}
\author[f,k]{Jason D. Rhodes}
\author[a,b,c,d]{L. Javier Romualdez}
\author[d]{J\"urgen Schmoll}
\author[h]{Sut-Ieng Tam}

\affil[a]{University of Toronto Institute for Aerospace Studies (UTIAS), 4925 Dufferin Street, Toronto, ON, Canada}
\affil[b]{Department of Physics, University of Toronto, 60 St. George Street, Toronto, ON, Canada}
\affil[c]{Dunlap Institute, University of Toronto, 50 St. George Street, Toronto, ON, Canada}
\affil[d]{Centre for Advanced Instrumentation (CfAI), Durham University, Science Laboratories, South Road, Durham, UK}
\affil[e]{Department of Physics, Princeton University, Washington Road, Princeton, NJ, USA}
\affil[f]{Jet Propulsion Laboratory (JPL), California Institute of Technology, 4800 Oak Grove Drive, Pasadena, CA, USA}
\affil[g]{Department of Astronomy/Steward Observatory, 933 North Cherry Avenue, Tucson, AZ 85721-0065, USA}
\affil[h]{Centre for Extragalactic Astrophysics, Durham University, South Road, Durham, UK}
\affil[i]{Department of Astronomy, University of Toronto, 50 St. George Street, Toronto, ON, Canada}
\affil[j]{Canadian Institute for Advanced Research, 661 University Ave., Suite 505, Toronto, ON, Canada}
\affil[k]{California Institute of Technology, 1201 East California Blvd, Pasadena, CA, USA} 

\authorinfo{Further author information: (Send correspondence to S.Redmond)\\E-mail: susan.redmond@mail.utoronto.ca, Telephone: 1 416 946 0946\\}

\begin{document} 
\maketitle
\begin{abstract}
Balloon-borne experiments present unique thermal design challenges, which are a combination of those present for both space and ground experiments.  Radiation and conduction are the predominant heat transfer mechanisms with convection effects being minimal and difficult to characterize at 35-40 km.  This greatly constrains the thermal design options and makes predicting flight thermal behaviour very difficult.  Due to the limited power available on long duration balloon flights, efficient heater control is an important factor in minimizing power consumption.  SuperBIT, or the Super-pressure Balloon-borne Imaging Telescope, aims to study weak gravitational lensing using a 0.5m modified Dall-Kirkham telescope capable of achieving 0.02" stability\cite{javierspie2016} and capturing deep exposures from visible to near UV wavelengths. To achieve the theoretical stratospheric diffraction-limited resolution of 0.25"\cite{javierphdthesis}, mirror deformation gradients must be kept to within 20 nm. The thermal environment must be stable on time scales of an hour and the thermal gradients on the telescope must be minimized.  During its 2018 test-flight, SuperBIT will implement two types of thermal parameter solvers: one for post-flight characterization and one for in-flight control.  The payload has 85 thermistors as well as pyranometers and far-infrared sensors which will be used post-flight to further understand heat transfer in the stratosphere.  This document describes the in-flight thermal control method, which predicts the thermal circuit of components and then auto-tunes the heater PID gains.  Preliminary ground testing shows the ability to control the components to within 0.01 K.

\end{abstract}

\keywords{balloon-borne, thermal control, wide field, visible-to-near-UV, stratosphere, auto-tune, PID gains}

\section{INTRODUCTION}
\label{sec:intro}  
\subsection{Weak gravitational lensing as a cosmological probe}\label{wlprobe}
SuperBIT's main science goal is to measure galaxy-galaxy weak gravitational lensing for up to 100 clusters.  This type of lensing uses galaxies at different redshifts; the distant galaxies as sources and a foreground cluster as the lens.  The SuperBIT telescope has been developed to have a $0.5^{\circ}$ field-of-view with a throughput above 80\% for the near-UV to near-infrared wavelengths to meet the weak gravitational lensing requirements.  Shape exposures are on the order of an hour followed by shorter exposures in the five photometry bands.  For this observing strategy it is critical that the thermal behaviour of the optics is extremely stable on hour timescales.  Oscillations of even a degree can have severe effects on the image quality and science goals. 

Weak gravitational lensing is sensitive to both statistical and instrument systematics.  Misalignment or over-constraining the optical surfaces can drastically affect the Point Spread Function (PSF).  It is important to note that SuperBIT is currently operating with an engineering telescope and the science instrument will have much higher quality optics.  Misalignment and over-constraining the optics can be induced by thermal deformations interacting poorly with the optics mounts. In addition to issues induced by isothermal contractions of the mirrors, thermal gradients cause non-uniformities which results in the PSF changing across the focal plane.  To account for this, SuperBIT dithers between exposures, which moves the image slightly on the focal plane further facilitating characterization of the systematics.  While post-processing techniques can help, systematic errors should be minimized from the start.
     
\subsection{Thermal control of SuperBIT optics in the stratosphere}\label{stratcont}

The stratosphere starts at an altitude of 10-18 km and transitions into the mesosphere at approximately 50 km \cite{atmosphere}.  The air pressure is small at 0.1 kPa and the temperature ranges from 222-278 K \cite{atmosphere2}.  An interesting characteristic of the stratosphere is that ambient temperature increases with increasing altitude due to absorbed UV radiation.  Due to this temperature gradient, there is no convection between stratospheric layers.  There are however winds within the stratospheric layers which move the balloon throughout the flight.  Also, throughout day-night cycles, temperature changes cause altitude variations due to pressure changes in the balloon.  This amplifies the temperature drop at night slightly as the balloon sinks to a cooler layer of the stratosphere.  To mediate this, it is possible to drop ballast making the payload lighter and thus causing the balloon to rise, but this is only done in practise when significant altitude changes occur.  Also, since balloon telescopes operate close to the Earth and their trajectory is uncontrolled, albedo and planetshine provide large variable heat loads.  The wavelength dependence of the incident energy combined with the uncertainty of the quantity hinders the loading prediction accuracy. Historically the conduction and convective effects to the residual atmosphere have been ignored but it is not clear that this is a suitable assumption.  For these reasons the stratosphere is a very harsh environment to operate in and difficult to characterize. 

\begin{figure}[h]
	\centering
	\includegraphics[scale=0.45]{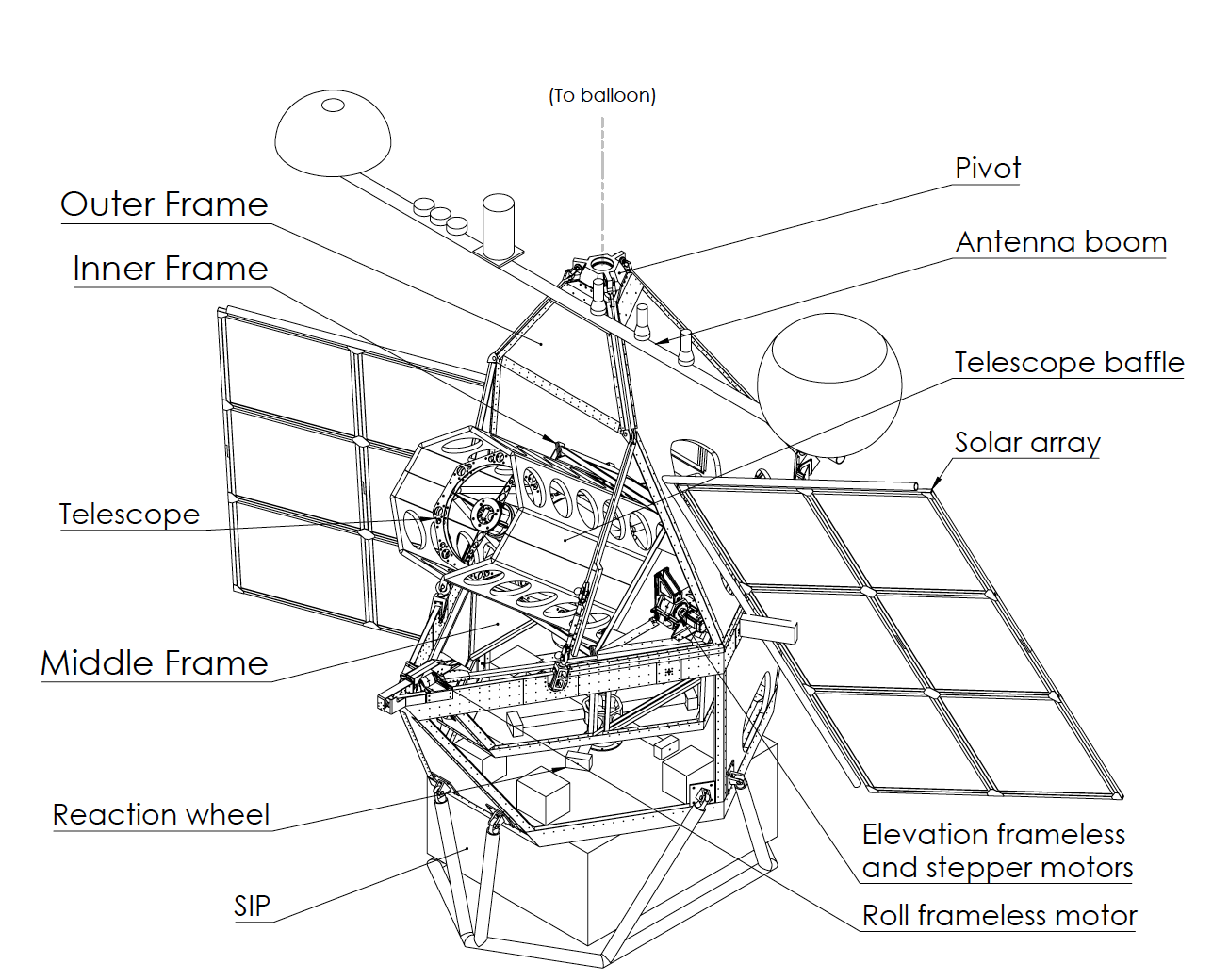}
	\captionsetup{justification=centering}
	\caption{SuperBIT \cite{javierphdthesis}}
	\label{fig:BIT} 
\end{figure}

The SuperBIT instrument is shown in Figure \ref{fig:BIT} and consists of three nested frames which allow for telescope control in roll, pitch, and yaw.  The coarse pointing system is used to stabilize the telescope to within 1" and the piezo-actuated tertiary mirror further improves it by two orders of magnitude to within 0.02" \cite{javierphdthesis}.  Fully assembled it stands three meters tall and weighs approximately one metric tonne.  For the three test flights a cost effective semi-custom telescope is used which is sufficient to determine pointing capabilities and provide a proof of concept imaging.  For the science flight a fully custom, high quality instrument has been designed which accounts for the unique SuperBIT requirements.  This telescope will be integrated and tested on the ground in late 2018.       
\begin{figure}[h]
	\centering
	\includegraphics[scale=1]{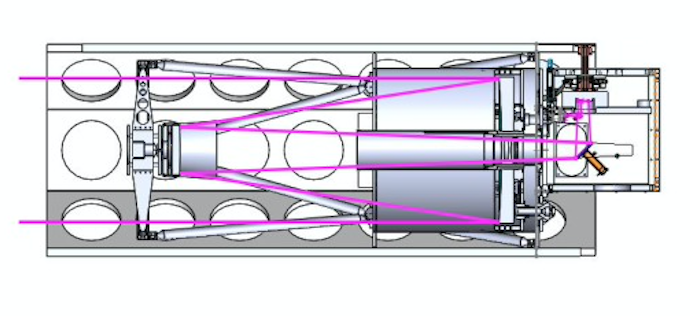}
	\captionsetup{justification=centering}
	\caption{SuperBIT Ray Trace Top View \cite{matcasca}}
	\label{fig:bitraytrace}
\end{figure}  

\begin{figure}[h]
	\centering
	\includegraphics[scale=0.4]{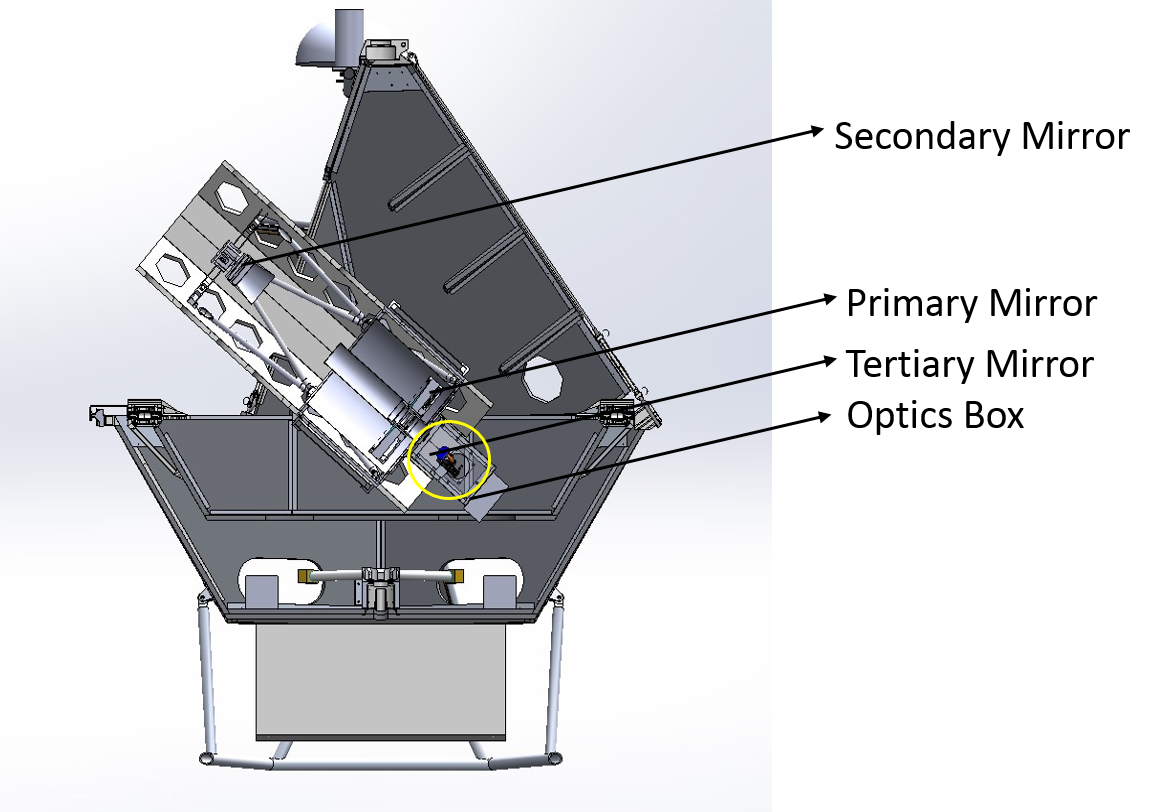}
	\captionsetup{justification=centering}
	\caption{Section View of SuperBIT}
	\label{fig:SectionBIT} 
\end{figure}

Thermal control of optical surfaces is a well known challenge in astronomy.  Static deformations on the order of 20 nm and varying deformations over time can result in misleading weak gravitational lensing measurements.  It is thus desired to house the instrument in a static, isothermal environment and to have a well designed telescope which allows the optics to contract freely at float.  SuperBIT has a 1.9m long temperature controlled baffle as shown in Figure \ref{fig:bitraytrace} to ensure the entire optics chain is in an isothermal environment.  The optics are controlled to 273 K for the test flights due to the thermal masses of the mirrors, sub-par mirror mounts, and steady state temperature of the tertiary mirrors.  The large thermal masses of the primary and secondary mirrors require a setpoint temperature which allows them to stabilize post-launch in a reasonable amount of time.  It was predicted that the sub-optimal design of the engineering telescope would excessively deform the mirrors if they were permitted to cool to the ambient 233 K.  The science telescope will be installed after the 2018 test flight making this concern obsolete.  Lastly, the optics box housing the tertiary mirror, shown in Figure \ref{fig:SectionBIT}, nominally reaches a temperature 273 K when the baffle heaters are off.  It is beneficial to have all optics at the same temperature so 273 K was a logical setpoint.  The secondary mirror is enclosed in an aluminum housing reducing the effectiveness of the temperature controlled baffle.  A heater is attached to a radiator plate located behind the secondary mirror which provides additional control while avoiding gradients across the optical surface.  

\subsection{Super pressure balloon payload heater configurations}

SuperBIT's science flight is designed for 100 days on a Super Pressure Balloon (SPB) out of Wanaka, NZ in 2020.  Due to the extreme day and night conditions during SPB flights, many of SuperBIT's electrical and mechanical components require heaters.  As of the 2016 test flight there are a total of 30 heaters on the payload and each heater is controlled using a standard proportional-integral-derivative (PID) control loop.  To minimize power and increase thermal stability, it is desired to have individual PID gains for each heater.  As previously discussed, the stratospheric environment is not well characterized so PID gains tuned on the ground are not necessarily well suited to float conditions.  Also, for components with a large thermal mass, tuning gains can take upwards of an hour.  Manually tuning 30 heaters is thus time consuming and impractical for a test flight scenario where time is limited.  For this reason an auto-tuning 
feature has been implemented.  Each heater-themistor-component system is modelled as an equivalent thermal resistor-capacitor (RC) circuit, and a least-squares optimizer is used to find the PID gains which minimize overshoot and settling time.  This is done using Matlab's sequential quadratic programming function \texttt{fmincon} described in Section \ref{sec:matopt}.  This function is relatively robust with respect to local minima and invalid guesses.
 
\section{METHODS}
\subsection{Heater thermal circuit}

\begin{figure}[H]
	\centering
	\begin{subfigure}[b]{0.4\textwidth}
		\includegraphics[scale = 0.5]{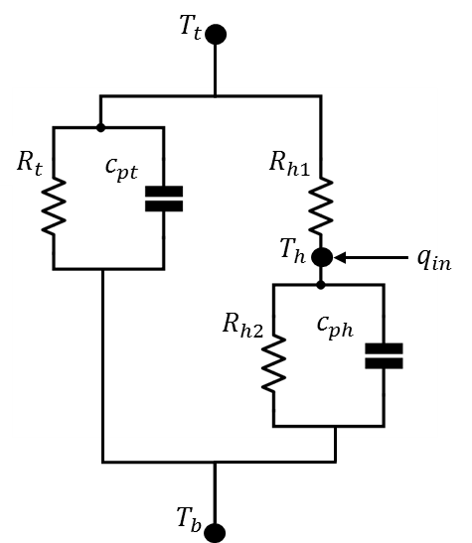}
		\centering
		\captionsetup{justification=centering}
		\caption{Full Circuit}
		\label{fig:qcirc1}
	\end{subfigure}
	\begin{subfigure}[b]{0.4\textwidth}
		\includegraphics[scale = 0.5]{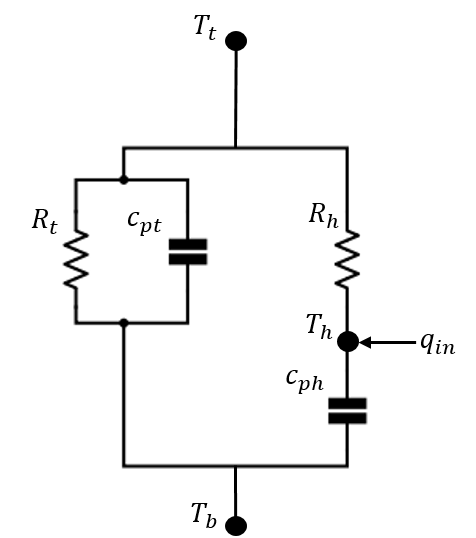}
		\centering
		\captionsetup{justification=centering}
		\caption{Approximated Circuit}
		\label{fig:qcirc2}
	\end{subfigure}
	\caption{Thermal Circuit for Thermistor-Heater-Bath System}
	\label{fig:qcirc}
\end{figure}

Each thermistor is assumed to interact with both the heater and a thermal bath.  The bath temperature is assumed to be constant and can often be approximated as the temperature of the gondola near the component in question. 
The full thermal circuit, as shown in Figure \ref{fig:qcirc1}, has both a capacitor and resistor between each node and the bath as well as a resistor between the two nodes.  In this figure $t$ indicates thermistor, $h$ indicates heater, and $b$ indicates bath.  Due to the long thermal link between the heater and the bath, the resistance between the heater and the bath will be very high.  This will push the net resistance along that chain toward $R_{h1}$.  Figure \ref{fig:qcirc2} shows the reduced circuit using the above assumption.   

For the reduced circuit (Figure \ref{fig:qcirc2}) the ordinary differential equation (ODE) can be written as:

\begin{equation} \label{eq:odeheater}
\begin{Bmatrix}
\dot{T_{h}}  \\
\dot{T_{t}}  \\
\end{Bmatrix} = 
\begin{bmatrix}
c_{ph} & 0\\
0 & c_{pt}\\
\end{bmatrix}^{-1}
\left[
\begin{bmatrix}
\frac{-1}{R_h} & \frac{1}{R_h}\\
\frac{1}{R_h} & -\left(\frac{1}{R_h} + \frac{1}{R_t}\right)\\
\end{bmatrix}
\begin{Bmatrix}
T_{h} \\
T_{t}  \\
\end{Bmatrix}
+
\begin{bmatrix}
1 & 0\\
0 & \frac{1}{R_t}\\
\end{bmatrix}
\begin{Bmatrix}
q_{in} \\
T_{b}  \\
\end{Bmatrix}
\right]
\end{equation}

Or

\begin{equation}
\dot{T} = \bm{C}^{-1}\left[\bm{R}T + \bm{P}Q\right]
\end{equation}

Where $T$ is the temperature array, $\bm{C}$ is the capacitance matrix, $\bm{R}$ is the resistance matrix, and $\bm{P}Q$ is the heat into the system from both the heater and the bath.  This ODE is solved using the Runge-Kutta or RK4\cite{rk4} approach with a user defined time step of $h$.  The $q_{in}$ profile can either be obtained directly from recorded data, or calculated using the same PID loop which controls the heaters.

\subsection{Heater control loop}

The heaters are controlled using a simple PID control loop.  First, the error ($E$), integrated error ($IE$), and change in error ($dE$) are computed between the setpoint and current temperature.  Each error measurement is weighted by a gain, where $K_p$ is the proportional gain, $K_i$ is the integral gain, and $K_d$ is the derivative gain.  The result is then used to determine the percent of total power applied by the heater.  The optimization loop uses a time step $h$ which can be different from the 100Hz flight PID loop calculation rate.  This requires a factor of $100h$ (the ratio between the flight and optimization time steps) to be incorporated into any term dependant on the time step.  To ensure long term stability, the current value of the error is de-weighted when calculating the integral.  This ensures that sporadic temperature or readout spikes do not drive the power request as this can result in a negatively damped system. The current temperature error receives a weight of $a$ and the integrated error from the previous time step receives a weight of $b$  which are empirically calibrated values: \\ 
$a = 5\times10^{-6}(100h)$\\
$b = 1-a$ \\

Mathematically the PID loop is:      

\begin{eqnarray}
E_i &=& T_{sp} - T_{ti} \label{eq:tsperr}\\
dE_i &=& -\left(E_i - E_{i-1}\right)\frac{1}{100h}\\
IE_i &=& 
\begin{cases}
aE_i + bIE_{i-1}, &IE_i > 0 \\
0, &IE_i < 0 \\
\end{cases}\\
P_{out_i} &=& 
\begin{cases}
0, &P_{out} < 0 \\
100, &P_{out} > 100 \\
K_pE_i + K_d dE_i + K_i IE_i, &0 < P_{out} < 100  \\
\end{cases}
\end{eqnarray}

Where $T_{sp}$ is the set point temperature, $T_{t}$ is the thermistor temperature, $P_{out}$ is the percent power request sent to the heater, and $i$ is the time step index.  

\subsection{Matlab optimizer} \label{sec:matopt}

To optimize the system, Matlab's \texttt{fmincon} function was chosen which uses Sequential Quadratic Programming (SQP) to minimize the objective function with respect to the constraints.  The ability to apply constraints to the optimization parameters is an important feature as the circuit parameters must be physical.  This is a gradient based optimizer where the gradients assist in choosing the search direction and updating the parameters within \texttt{fmincon}.  Benefits of this approach include its ability to find global minima, strictly respect the bounds, and to recover from steps which violate the bounds \cite{sqp}.  

The general \texttt{fmincon} form is: 

\begin{equation}
\min\limits_{x} J(x)= 
\begin{dcases}
\bm{A}x \leq b\\
\bm{A_{eq}}x = b_{eq}\\
L_b \leq x \leq U_b
\end{dcases}
\label{eq:fmincon}
\end{equation}

Here $x$ is the vector containing the unknown parameters and $J$ is the objective function.  The $\bm{A}$ matrices and $b$ arrays allow for equality and inequality constraints to be enforced between the elements.  Lower and upper bounds can also be applied to the parameters.  For this problem, both the circuit parameters and PID gains are constrained to be larger than zero.  The  objective function depends on the particular problem but is generally closely related to the error between the predicted temperature profile and the desired temperature profile.   

\subsection{Heater gain auto-tune methodology}

The general process for auto-tuning the heater PID gains is as follows: 

\textbf{Step 1:} The component heater is turned on to 100\% until the thermistor sees an increase in temperature at which point it is turned off.  Temperature and heater power data ($T_{dat}$ \& $Q_{dat}$) is recorded until the component returns to its original temperature.  The magnitude of the increase is chosen to be $\sim$2 K for components with a large thermal mass and $\sim$4 K for components with a small thermal mass.  These $\Delta T$ values were empirically determined to produce an adequate curve for the optimizer to fit as it works best using a fine time step ($\sim 0.2$ s) and a shorter total time.    \\
Output: $T_{dat} , Q_{dat}$\\

\textbf{Step 2:} The thermal circuit parameters and optimal PID gains are determined.\\	

\textit{Step 2a:} The solver is run using the recorded heater profile to determine thermal resistances and capacitances of the system.\\
\begin{figure}[H]
	\centering
	\includegraphics[scale=0.5]{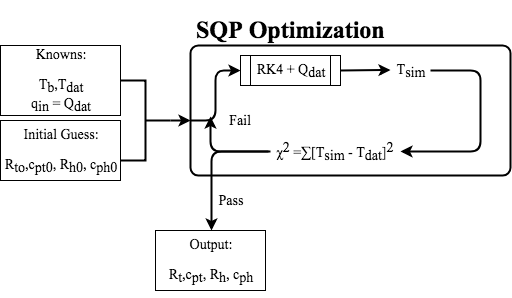}
	\captionsetup{justification=centering}
	\caption{Thermal Circuit Parameter Solver Algorithm}
	\label{fig:heatcontloop1} 
\end{figure}
Output: $R_t, c_{pt} , R_h , c_{ph}$\\

For this step the recorded temperature response of the component, $T_{dat}$, is used as the desired temperature profile.  Initial guesses for the circuit parameters are fed into Matlab's \texttt{fmincon} function.  The RK4 approach is used to determine the temperature profile for the current parameter values using the known heater pulse, $Q_{dat}$, as the heat input.  This produces the profile $T_{sim}$ which is then compared to $T_{dat}$.  The sum of the squared error is then minimized by adjusting the thermal circuit parameters until a best fit is reached.

\begin{figure}[H]
	\centering
	\includegraphics[scale=0.47]{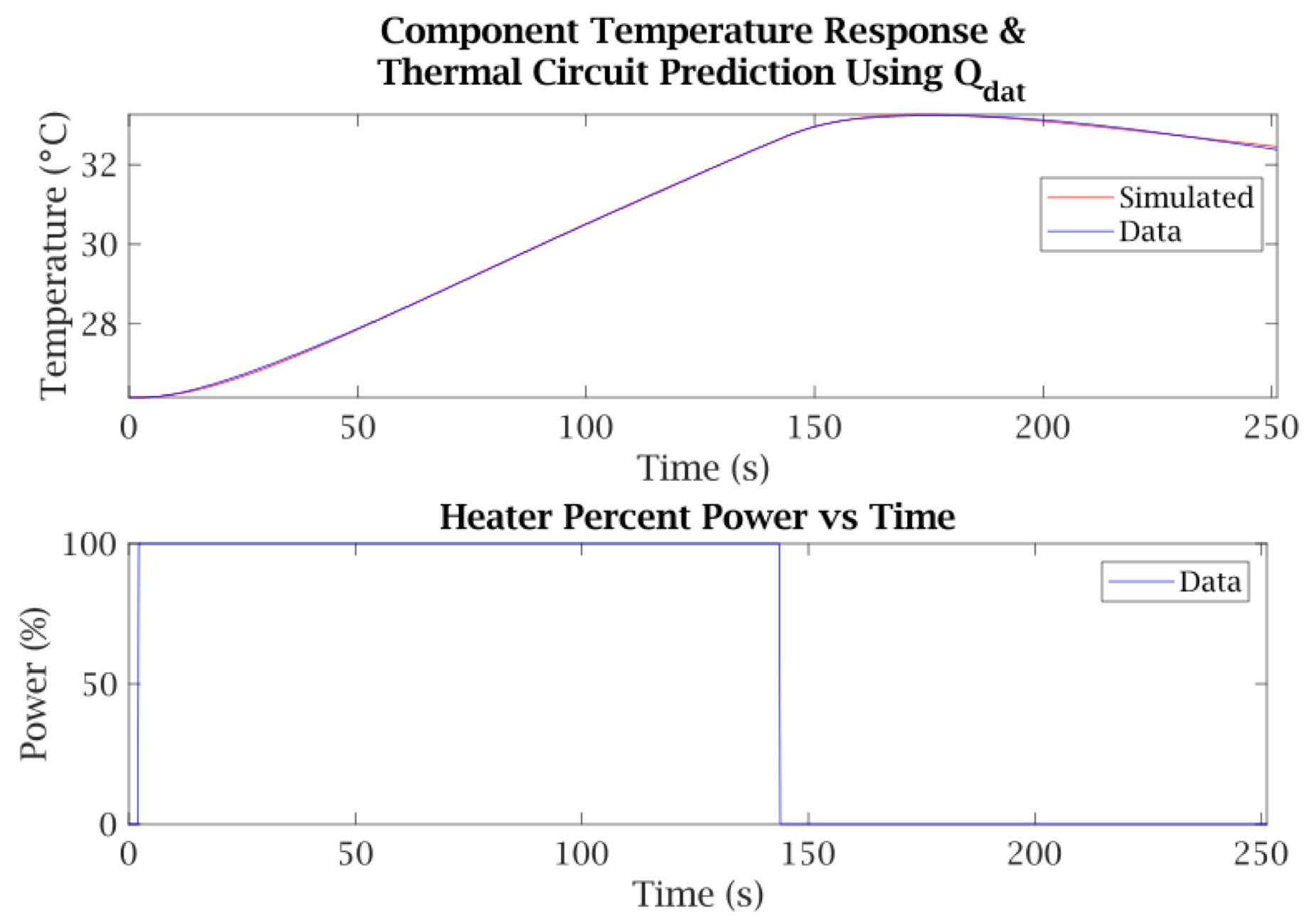}
	\captionsetup{justification=centering}
	\caption{Thermal Circuit Model - Component Comparison for Step Input}
	\label{fig:thermcircopt} 
\end{figure}

Figure \ref{fig:thermcircopt} shows the thermal response to the heater step input in blue for one of SuperBIT's telescope baffle panels.  The red curve shows the predicted behaviour using the circuit parameters produced by the optimizer.  Both the profile and the magnitude of the simulated curve align with the data showing the circuit is a good match for the system.  

\textit{Step 2b:} Using above RC values, the solver is run with the full PID loop to determine optimal PID gains.  The start temperature ($T_0 = T(t=0)$) and setpoint temperature ($T_{sp}$) are chosen such that $T_{sp} - T_0$ follows the same guidelines as described in \textit{Step 1}.  The response characteristics used to determine the quality of the gains are\cite{pidmat}:\\
$P_o$ - Percent overshoot: percent above $T_{sp}$ thermistor reaches on the first oscillation\\
$t_{rise}$ - Rise time: time to rise from 10\% to 90\% of $T_{sp} - T_0$ \\
$t_{set}$ - Settling time: time for the amplitude of the error to reduce to 0.1 K \\
$E_{ss}$ - Steady state error: maximum error after $t_{set}$ is reached \\

Within the optimizer, each parameter is also weighted by a factor $w$.  For this application the system behaviour is desired to be slightly under-damped.  For this reason the weights in decreasing order are the percent overshoot, steady state error, settling time, and rise time.  It was found that if the rise time was significantly weighted it resulted in a large overshoot and extended settling time, neither of which are desired.  Also, due to the modified integral term in the PID loop, the steady state error must be weighted heavily in order for the system to reach the setpoint temperature.  This weighting scheme produces PID gains which reflect the desired temperature profile for each component.        

\begin{figure}[H]
	\centering
	\includegraphics[scale=0.5]{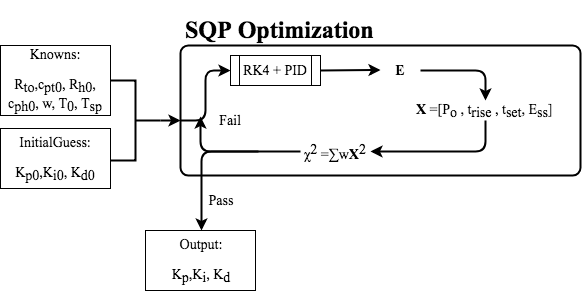}
	\captionsetup{justification=centering}
	\caption{PID Gain Solver Algorithm}
	\label{fig:heatcontloop2} 
\end{figure}
Output: $K_p , K_i , K_d$\\

This process is very similar to Step 2a except the thermal circuit parameters are known and the control loop parameters are unknown.  The thermal circuit parameters, user defined setpoint, and initial gain guesses are fed into \texttt{fmincon} but instead of using $Q_{dat}$, the flight PID loop is implemented when solving the ODE.  The error between the setpoint and thermistor (Equation \ref{eq:tsperr}) is used to calculate the response characteristics listed above.  The sum of the weighted response characteristics is minimized to find the optimal PID gains.  This loop is sensitive to the initial guess and a poor guess could result in a sub-optimal solution.  For this reason the optimizer is run with a number of different initial guesses and the gains which produce the lowest $\chi ^2$ are used.  \\

\textbf{Step 3:} The optimized PID gains are verified. \\

\textit{Step 3a:} The PID gains are used to control component to a given temperature.\\

\textit{Step 3b:} The data is compared to the predicted performance for same setpoint; note that this is a model verification step only.\\

\begin{figure}[H]
	\centering
	\includegraphics[scale=0.43]{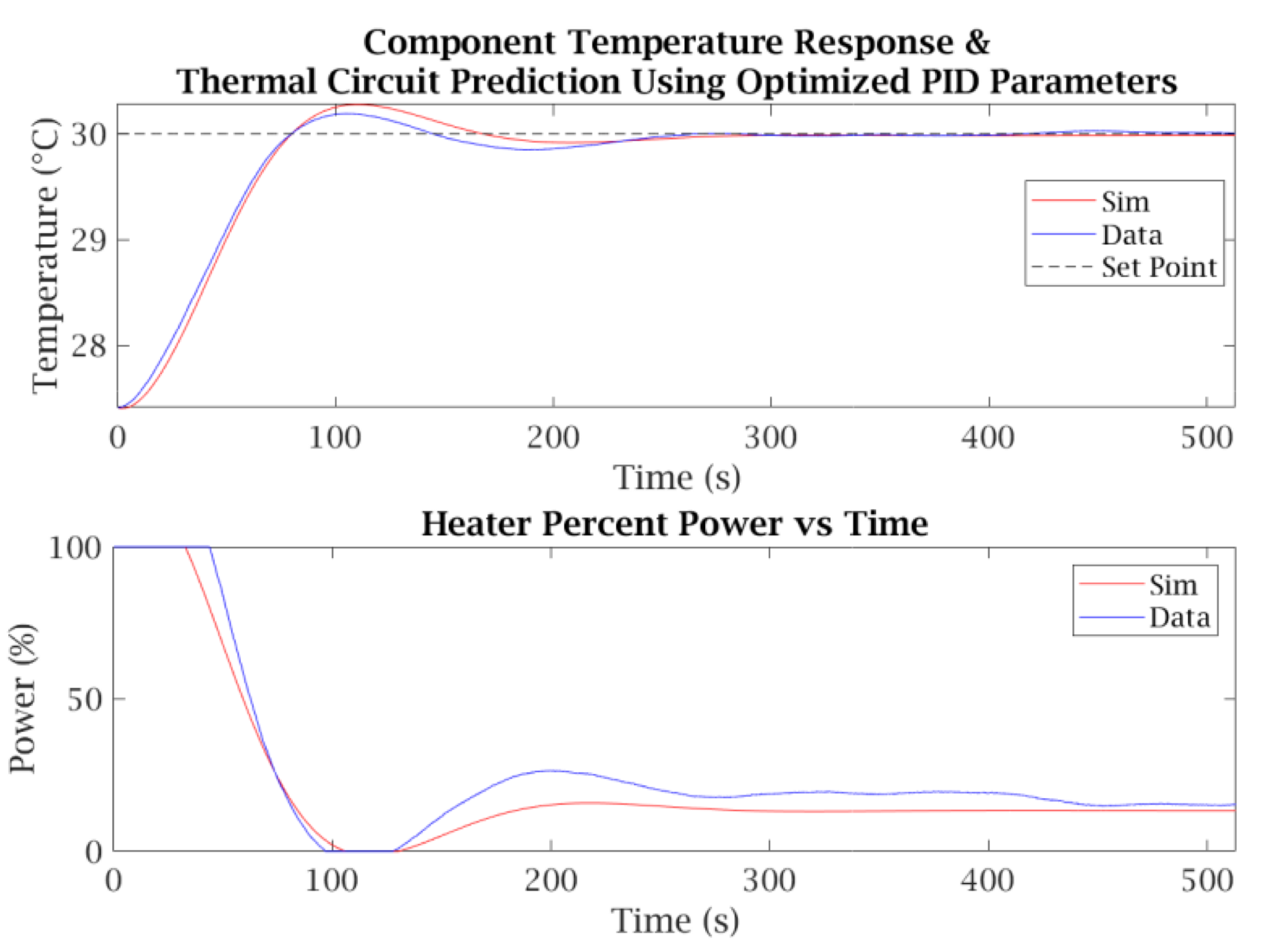}
	\captionsetup{justification=centering}
	\caption{Thermal Circuit Model - Component Comparison Using Optimized PID Gains}
	\label{fig:heatergaincomp} 
\end{figure}

Figure \ref{fig:heatergaincomp} shows the result of this process for one of the telescope baffle panels.  As with Figure \ref{fig:thermcircopt}, blue is the data, red is the predicted behaviour, and cyan is the setpoint.  The error is slightly larger than Figure \ref{fig:thermcircopt} but the profiles are well aligned.  The error between the two temperature curves also decreases with time and both curves approach the setpoint within a reasonable amount of time demonstrating the effectiveness of this process.      
   
This will become an automated process which will run steps one through three and then send the updated gains to the payload.  This allows dynamic updating throughout the flight as the environment changes. 

\section{FLIGHTS AND IMPLEMENTATION}

SuperBIT has completed two engineering test flights. The first was in 2015 with the Centre National d'\'Etudes Spatiales (CNES) and Canadian Space Agency (CSA) out of Timmins, ON and the second in 2016 with the Columbia Scientific Balloon Facility (CSBF-NASA) out of Palestine, TX.  There were no thermal concerns during either flight but since they were engineering flights, the requirements were more relaxed.  The upcoming test flight will be the final proof of concept and will collect data for weak gravitational lensing analysis on a cluster provided the optics of the engineering telescope are of sufficient quality.  The quality is uncertain as until recently, pointing capabilities were the limiting factor for image quality making it difficult to quantify telescope characteristics.  As the next flight has better defined science objectives, the thermal performance goals are much more ambitious.  The heater auto-tuning feature is intended to keep the thermal environment stable and will be rigorously tested throughout the flight.  Additionally, an updated heat pipe cooling arrangement has been implemented on both the science camera and focal plane star camera to minimize the dark current which reduces image quality.  The optics box has also been modified to provide additional thermal shielding between it and the primary mirror.  A total of 10 thermistors on the primary mirror and six along the telescope truss structure will be used to characterize and quantify the thermal loading on the telescope due to nearby electronics prior to the SPB flight from New Zealand in 2020.  These will help determine the quality of the thermal design currently implemented and provide guidance for the science telescope thermal design.  In addition to the thermal performance goals, this flight is an opportunity to characterize the stratospheric environment.  A total of 85 thermistors are being flown to enable thorough post-flight analysis between the predicted and recorded thermal behaviour of the entire payload. 

\section{CONCLUSIONS}

Due its unique operating conditions and capabilities, SuperBIT enables optical to near-UV imaging for a fraction of the time and cost of a space telescope while retaining image quality.  Using it for weak gravitational lensing measurements introduces extremely strict requirements on the thermal environment, optical performance, and pointing system.  The combination of the large number of heaters and the time associated with tuning each one makes it impossible to have individual PID gains for a test flight and cumbersome for a mission flight.  The heater auto-tuning feature will assist in maintaining a stable thermal environment throughout all future flights.

\acknowledgments 
 
SuperBIT is supported in Canada, via the Natural Sciences and Engineering Research Council (NSERC), in the USA via NASA award NNX16AF65G, and in the UK via the Royal Society and Durham University. Part of the research was carried out at the Jet Propulsion Laboratory (JPL), California Institute of Technology (Caltech), under a contract with NASA.

\bibliography{report} 
\bibliographystyle{spiebib} 

\end{document}